\documentclass[aps,amsmath,amssymb,nofootinbib,superscriptaddress,showpacs,floatfix,prl,twocolumn]{revtex4-1}

\usepackage{latexsym}
\usepackage{graphicx}
\usepackage{times,psfrag,subfigure}
\usepackage{enumerate}
\usepackage{amsmath}
\usepackage{dcolumn}
\usepackage{bm,bbm}
\usepackage{color}
\usepackage{latexsym,amsmath,amssymb,bm,euscript}
\usepackage{textcomp}
\usepackage{tabularx}
\usepackage{setspace}
\usepackage{ctable}
\usepackage{sidecap}
\usepackage{placeins}
\usepackage{threeparttable}
\usepackage{multirow}
\usepackage[utf8]{inputenc}
\usepackage[english, ngerman]{babel}
\def\b0{{\bm0}}

\newcommand{\beq}{\begin{equation}}
\newcommand{\eeq}{\end{equation}}
\newcommand{\beqarray}{\begin{eqnarray}}
\newcommand{\eeqarray}{\end{eqnarray}}

\graphicspath{{./Images/}}

\begin{document}
\selectlanguage{english}
\allowdisplaybreaks

\title{Tunable Weyl and Dirac states in the nonsymmorphic compound $\rm\mathbf{CeSbTe}$}

\date{\today}

\author{Leslie M. Schoop}
\email{l.schoop@fkf.mpg.de}
\affiliation{Max-Planck-Institut f\"ur Festk\"orperforschung, Heisenbergstra\ss e 1, D-70569 Stuttgart, Germany}

\author{Andreas Topp}
\affiliation{Max-Planck-Institut f\"ur Festk\"orperforschung, Heisenbergstra\ss e 1, D-70569 Stuttgart, Germany}

\author{Judith Lippmann}
\affiliation{Max-Planck-Institut f\"ur Festk\"orperforschung, Heisenbergstra\ss e 1, D-70569 Stuttgart, Germany}

\author{Fabio Orlandi}
\affiliation{Science and Technology Facilities Council Rutherford Appleton, ISIS neutron pulsed facility, OX11 0QX, United Kingdom}

\author{Lukas Müchler}
\affiliation{Department of Chemistry, Princeton University, Princeton, New Jersey 08544, USA}

\author{Maia G. Vergniory}
\affiliation{Donostia International Physics Center, Paseo Manuel de Lardizabal 4, 20018 Donostia-San Sebastian, Spain}

\author{Yan Sun}
\affiliation{Max-Planck-Institut f\"ur Chemische Physik fester Stoffe, 01187 Dresden, Germany}

\author{Andreas W. Rost}
\affiliation{Max-Planck-Institut f\"ur Festk\"orperforschung, Heisenbergstra\ss e 1, D-70569 Stuttgart, Germany}
\affiliation{Physikalisches Institut, Universit\"at Stuttgart, Pfaffenwaldring 57, 70569 Stuttgart, Germany}

\author{Viola Duppel}
\affiliation{Max-Planck-Institut f\"ur Festk\"orperforschung, Heisenbergstra\ss e 1, D-70569 Stuttgart, Germany}

\author{Maxim Krivenkov}
\affiliation{Helmholtz-Zentrum Berlin f\"ur Materialien und Energie, Elektronenspeicherring BESSY II, Albert-Einstein-Stra\ss e 15, 12489 Berlin, Germany}

\author{Shweta Sheoran}
\affiliation{Max-Planck-Institut f\"ur Festk\"orperforschung, Heisenbergstra\ss e 1, D-70569 Stuttgart, Germany}

\author{Pascal Manuel}
\affiliation{Science and Technology Facilities Council Rutherford Appleton, ISIS neutron pulsed facility, OX11 0QX, United Kingdom}

\author{Andrei Varykhalov}
\affiliation{Helmholtz-Zentrum Berlin f\"ur Materialien und Energie, Elektronenspeicherring BESSY II, Albert-Einstein-Stra\ss e 15, 12489 Berlin, Germany}

\author{Binghai Yan}
\affiliation{Department of Condensed Matter Physics, Weizmann Institute of Science, Rehovot 76100, Israel}

\author{Reinhard K. Kremer}
\affiliation{Max-Planck-Institut f\"ur Festk\"orperforschung, Heisenbergstra\ss e 1, D-70569 Stuttgart, Germany}

\author{Christian R. Ast}
\affiliation{Max-Planck-Institut f\"ur Festk\"orperforschung, Heisenbergstra\ss e 1, D-70569 Stuttgart, Germany}

\author{Bettina V. Lotsch}
\email{b.lotsch@fkf.mpg.de}
\affiliation{Max-Planck-Institut f\"ur Festk\"orperforschung, Heisenbergstra\ss e 1, D-70569 Stuttgart, Germany}
\affiliation{Department of Chemistry, Ludwig-Maximilians-Universit\"at M\"unchen, Butenandtstr. 5-13, 81377 M\"unchen, Germany}
\affiliation{Nanosystems Initiative Munich (NIM) \& Center for Nanoscience, Schellingstr. 4, 80799 M\"unchen, Germany}

\begin{abstract}
Recent interest in topological semimetals has lead to the proposal of many new topological phases that can be realized in real materials. Next to Dirac and Weyl systems, these include more exotic phases based on manifold band degeneracies in the bulk electronic structure. The exotic states in topological semimetals are usually protected by some sort of crystal symmetry and the introduction of magnetic order can influence these states by breaking time reversal symmetry. Here we show that we can realize a rich variety of different topological semimetal states in a single material, $\rm CeSbTe$. This compound can exhibit different types of magnetic order that can be accessed easily by applying a small field. It allows, therefore, for tuning the electronic structure and can drive it through a manifold of topologically distinct phases, such as the first nonsymmorphic magnetic topological material with an eight-fold band crossing at a high symmetry point. Our experimental results are backed by a full magnetic group theory analysis and {\it ab initio} calculations. This discovery introduces a realistic and promising platform for studying the interplay of magnetism and topology.

\end{abstract} 

\date{\today}

\pacs{71.20.Eh, 61.05.fm, 75.50.Ee, 81.05.Bx}

\maketitle

\section{Introduction}
In recent years, the field of topological semimetals has flourished due to the discovery of many materials that exhibit highly exotic physical properties that are a result of massless quasiparticles, which dominate the transport properties \cite{wang2012dirac,wang2013three,gibson2015three,neupane2014observation,vafek2014dirac,ali2014crystal,borisenko2014experimental,
liu2014discovery,lv2015observation,soluyanov2015new,burkov2011topological,moll2016transport}. For example, 3D Dirac semimetals (3D DSMs) and Weyl semimetals (WSMs) have been shown to host unusual electronic transport properties, evident by their ultra high carrier mobility, their extremely large magnetoresistance and indications of the chiral anomaly \cite{liang2015ultrahigh,li2015negative,
xiong_evidence_2015,huang2015observation,du2015unsaturated,ali2014large,
ali2015correlation,Shekhar2015large,lv2016extremely,ali2016butterfly}. The current vast progress in the field draws the prospect of developing new spintronic devices based on topological materials closer \cite{vsmejkal2017route,sun2016strong}. In order to overcome existing limitations, such as the necessity of high magnetic fields to access several topological effects, new and improved materials are of high demand. Especially materials that exhibit long range magnetic order in combination with topologically non trivial band structures, are sparse. However, it is exactly these type of materials that may solve present challenges for spintronic applications \cite{vsmejkal2017electric}.

3D DSMs exhibit a four-fold degenerate point in their electronic structure, which is a result of a crossing of two doubly degenerate bands. If inversion symmetry (IS) or time reversal symmetry (TRS) is broken in a 3D DSM, the doubly degenerate bands become spin split, resulting in double degenerate band crossings called Weyl nodes \cite{wan2011topological}. While, through IS breaking, WSMs have been established for some time \cite{xu2015discovery,lv2015observation,xu2015discovery2,xu2015observation,weng2015weyl,
sun2015prediction,wang2016mote,nakayama2017band}, there are only very few examples of TRS breaking WSMs, examples being $\rm GdPtBi$ or cobalt based Heusler alloys \cite{hirschberger2016chiral,wang2016time,borisenko2015time}. Antiferromagnetic DSMs are even more rarely found \cite{tang2016dirac}.
Contrary to most 3D DSMs and WSMs, where the band degeneracy is dependent on the orbital character of the bands, materials that crystallize in nonsymmorphic space groups exhibit band degeneracies that are imposed by the space group symmetry \cite{young2012dirac,young2015dirac,bradlyn2016beyond,wieder2015double,wang2016hourglass}.
These degeneracies are guaranteed to be present if a nonsymmorphic symmetry operation relates atoms in the crystal structure. This concept has been used to predict DSMs as well as completely new quasi-particles (the so-called new fermions), beyond Dirac, Weyl or Majorana fermions \cite{bradlyn2016beyond,wieder2015double}. These, yet to be discovered, new fermions are not constrained by Poincar$\mathrm{\acute{e}}$ symmetry, they only exist in the solid state, and have no high energy counterparts \cite{bradlyn2016beyond}. Their unusual Landau level structure distinguishes them from normal Weyl points. The concept of creating such new quasi-particles is based on the possibility of achieving three-, six- or eight-fold band degeneracies in nonsymmorphic space groups. But whether these degeneracies appear at the Fermi level depends on the electron filling, which complicates the search for such topological semimetals \cite{watanabe2015filling,watanabe2016filling,topp2016non}. Three-fold degeneracies can also appear coincidentally in materials with only symmorphic symmetries \cite{lv2016experimental,weng2016topological}. So far, only very few materials have been experimentally shown to exhibit degeneracies protected by nonsymmorphic symmetry, examples being $\rm ZrSiS$ and $\rm ZrSiTe$ \cite{schoop2015dirac,topp2016non,chen2017dirac}. It has also not yet been studied experimentally how magnetism and thus TRS breaking affects the band degeneracies that result from nonsymmorphic symmetry, mostly due to a lack of such candidate materials. A magnetic nonsymmorphic compound could be a new type of magnetic Weyl semimetal that combines long range magnetic order with new quasi-particles.

In times where many topological materials have been realized, the challenge to introduce compounds that combine several different states in a single material and allow for switching between these different topological states, remains. Here we show that the compound $\rm CeSbTe$ exhibits Dirac and Weyl fermions, and that both types, nonsymmorphically protected and accidental band crossings, are present in the vicinity of the Fermi level. We also show that this material can exhibit more exotic three-fold and eight-fold degeneracies. While $\rm CeSbTe$ is centrosymmetric, TRS can be broken by applying a very small magnetic field. $\rm CeSbTe$ orders antiferromagnetically below $T_\mathrm{N} = 2.75\,$K at zero field but undergoes a metamagnetic transition to a fully polarized state under a small field of about 0.25\,T. This magnetically fully polarized state with ferromagnetic-like (FM-like) polarization will be referred to as FM phase within this manuscript. We show with detailed magnetic measurements combined with \textit{ab initio} calculations and ARPES measurements that this material allows for easy access to a plethora of different Dirac and Weyl states, of which some have not previously been realized in a real material. Using the recently implemented double group representations \cite{aroyo2011crystallography,aroyo2006bilbao1,aroyo2006bilbao2,
perez2015symmetry}, we performed a group theory analysis \cite{miller1967tables,aroyo2011crystallography,aroyo2006bilbao1,aroyo2006bilbao2,
perez2015symmetry,bradlyn2016beyond,bradley2010mathematical} that supports our \textit{ab initio} claims of the existence of protected Dirac and Weyls nodes, as well as three- and eight-fold crossings at high symmetry points or lines, thus presenting the first realization of a nonsymmorphic topological material that exhibits other than four-fold degeneracies. We also show that AFM order allows to access new fermion states in space groups that have previously not been considered.

\section{METHODS AND EXPERIMENTAL RESULTS}
Single crystals were grown with iodine vapor transport. The crystal structure was solved with single crystal x-ray diffraction (SXRD), neutron diffraction and electron diffraction (see supplemental information (SI) for details). Magnetic measurements were performed on a MPMS-XL and a PPMS equipped with a VSM option from Quantum Design. Specific heat measurements were performed on a PPMS from Quantum Design.
Powder neutron diffraction data were collected on the WISH instrument at ISIS, Harwell Oxford \cite{chapon2011wish} (see SI for details). For ARPES measurements crystals were cleaved and measured in ultra-high vacuum (low $10^{-10}$\,mbar range). Spectra were recorded with the $1^2$-ARPES experiment installed at the UE112-PGM2a beamline at BESSY-II in Berlin. The spectra were taken at room temperature. Density functional theory (DFT) calculations were performed using the augmented plane wave method as implemented in the Vienna ab-initio Simulation Package (VASP) \cite{kresse1996efficiency}. Due to the presence of Ce-f electrons, we have considered the exchange-correlation energy by the DFT+\textit{U} method with $U = 4$\,eV added to the Ce atoms \cite{dudarev1998electron}, reproducing the experimentally measured magnetic moment. All the electronic structures are calculated based on experimental lattice constants, including spin orbit coupling (SOC).

\begin{figure}[h]
  \centering
  \includegraphics[width = \columnwidth]{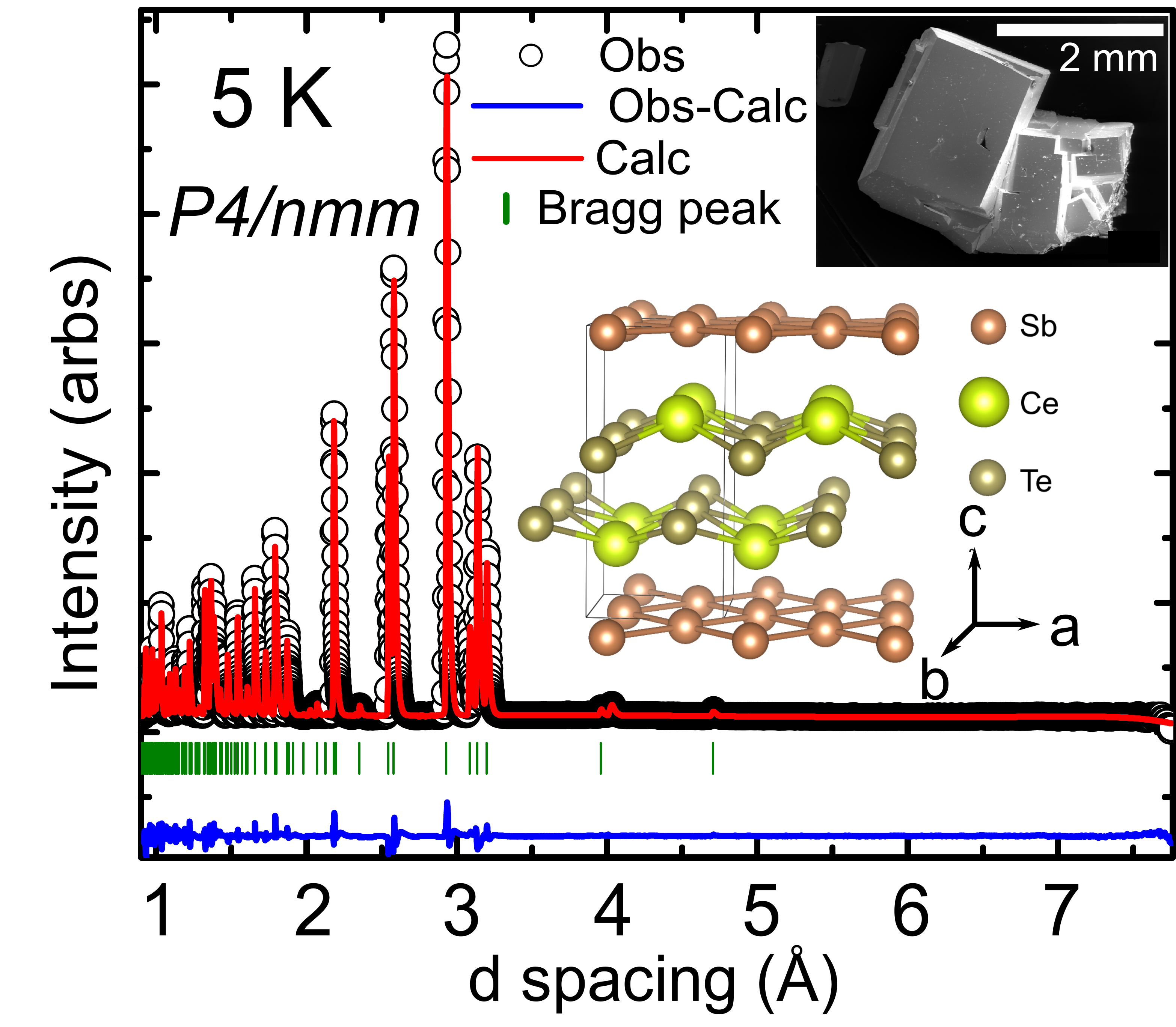}
  \caption{Refined neutron diffraction data taken at 5\,K. Small impurity peaks were excluded from the refinement. The upper inset shows an SEM image of a typical crystal of $\rm CeSbTe$, the lower inset shows a drawing of the crystal structure of $\rm CeSbTe$.}
  \label{str}
\end{figure}

Figure \ref{str} shows a drawing of the crystal structure of $\rm CeSbTe$; an image of a typical single crystal is shown in the inset. Contrary to a previous report \cite{wang2001structure}, which reported $\rm CeSbTe$ to crystallize in the orthorhombic space group \textit{Pnma} (No.\,62), we find it to crystallize in the tetragonal space group \textit{P4/nmm} (No.\,129), as evidenced by the refined neutron diffraction data (Fig.\,\ref{str}). SXRD and precession electron diffraction (PED) yielded the same result (see SI for details). With PED, we do not detect any of the additional reflections that would be expected for space group \textit{Pnma} (Fig.\,1 in the SI). Thus, $\rm CeSbTe$ is an isostrucutral and isoelectronic version of the nodal line semimetal $\rm ZrSiS$ \cite{schoop2015dirac,xu2015two}. The crystal structure is composed of $\rm CeTe$ bilayers that are separated by a square-net layer of Sb; the layers stack along \textit{c} (see inset of Fig.\,\ref{str}). The Ce atoms form a square network arrangement and the intralayer distance of Ce atoms is 4.37\,\AA, whereas the interlayer distance is 5.26\,\AA\, within a bilayer and 6.02\,\AA\, across a bilayer. $\rm ZrSiS$ and related compounds have been shown to exhibit a diamond shaped Dirac line node as well as four-fold degenerate nodes at the X, R, M and A points of the Brillouin zone (BZ) \cite{xu2015two,schoop2015dirac,topp2016non,takane2016dirac,hu2016evidence,lou2016emergence,
chen2017dirac,hu2017quantum}. The latter ones are a result of the nonsymmorphic glide planes in the space group \textit{P4/nmm}. While the Dirac line node is gapped by spin orbit coupling (SOC), the degeneracies that result from nonsymmorphic symmetry are not affected by SOC. The nonsymmorphic degeneracies can be above, below, or right at the Fermi level, depending on the \textit{c/a}-ratio of the lattice constants \cite{topp2016non}. 
For further information on the crystal structure and sample quality see HRTEM images with atomic resolution in the SI.

\begin{figure}[h]
  \centering
  \includegraphics[width = \columnwidth]{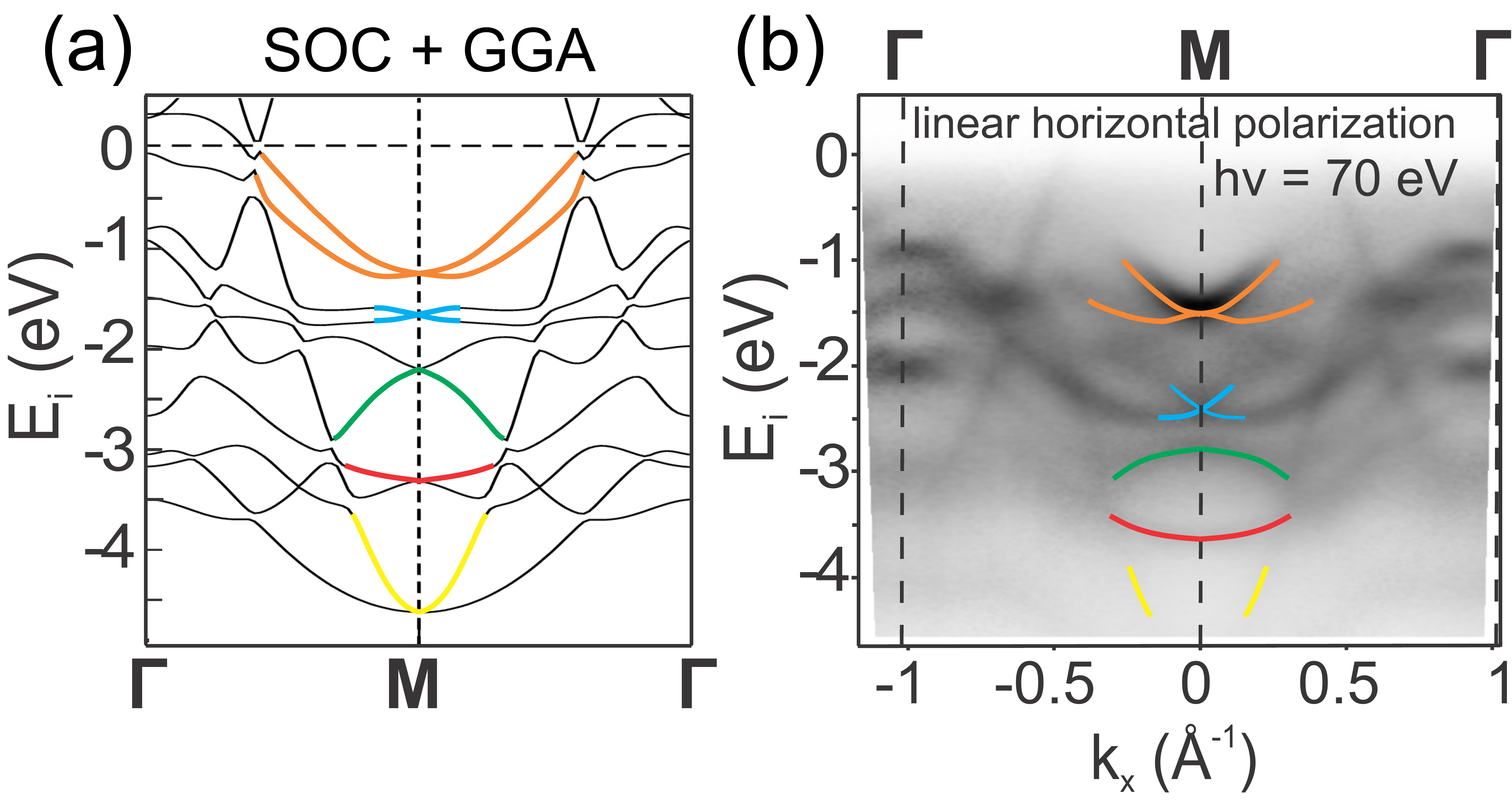}
  \caption{(a) Calculated bulk band structure plotted along $\Gamma$M$\Gamma$. (b) Dispersion along $\Gamma$M$\Gamma$, measured with ARPES, in comparison to the calculation. The crossings at M are sketched on top of the measured data. Except for the energy scaling, the measured band structure is in perfect agreement with the calculations.}
  \label{arpes}
\end{figure}

The clarification of the crystal structure allows us to calculate the electronic band structure, to find out if the system is not only structurally but also electronically similar to $\rm ZrSiS$ and related nonsymmorphic Dirac materials. An excerpt of the calculated electronic structure, without considering magnetic order, can be found in Fig.\,\ref{arpes}(a), showing the expected nonsymmorphically protected four-fold degeneracies at M, and a more detailed discussion will follow later in the paper.
Since DFT on \textit{f} electron systems has been proven to be challenging, we performed ARPES measurements on the paramagnetic phase (samples were measured at room temperature without a magnetic field), to confirm the predicted electronic structure. Fig.\,\ref{arpes}(b) displays the measured dispersion along $\Gamma$M$\Gamma$. Almost all bands are clearly resolved in the measurement. Some bulk bands are not visible, which we attribute to matrix element effects (see SI for more ARPES data along different high symmetry lines). Overall, the room temperature electronic structure of $\rm CeSbTe$ is in good agreement with the predicted paramagnetic band structure. We thus can be confident that further DFT predictions in this manuscript are sufficiently reliable. 

\begin{figure*}[h]
  \centering
  \includegraphics[width = \textwidth]{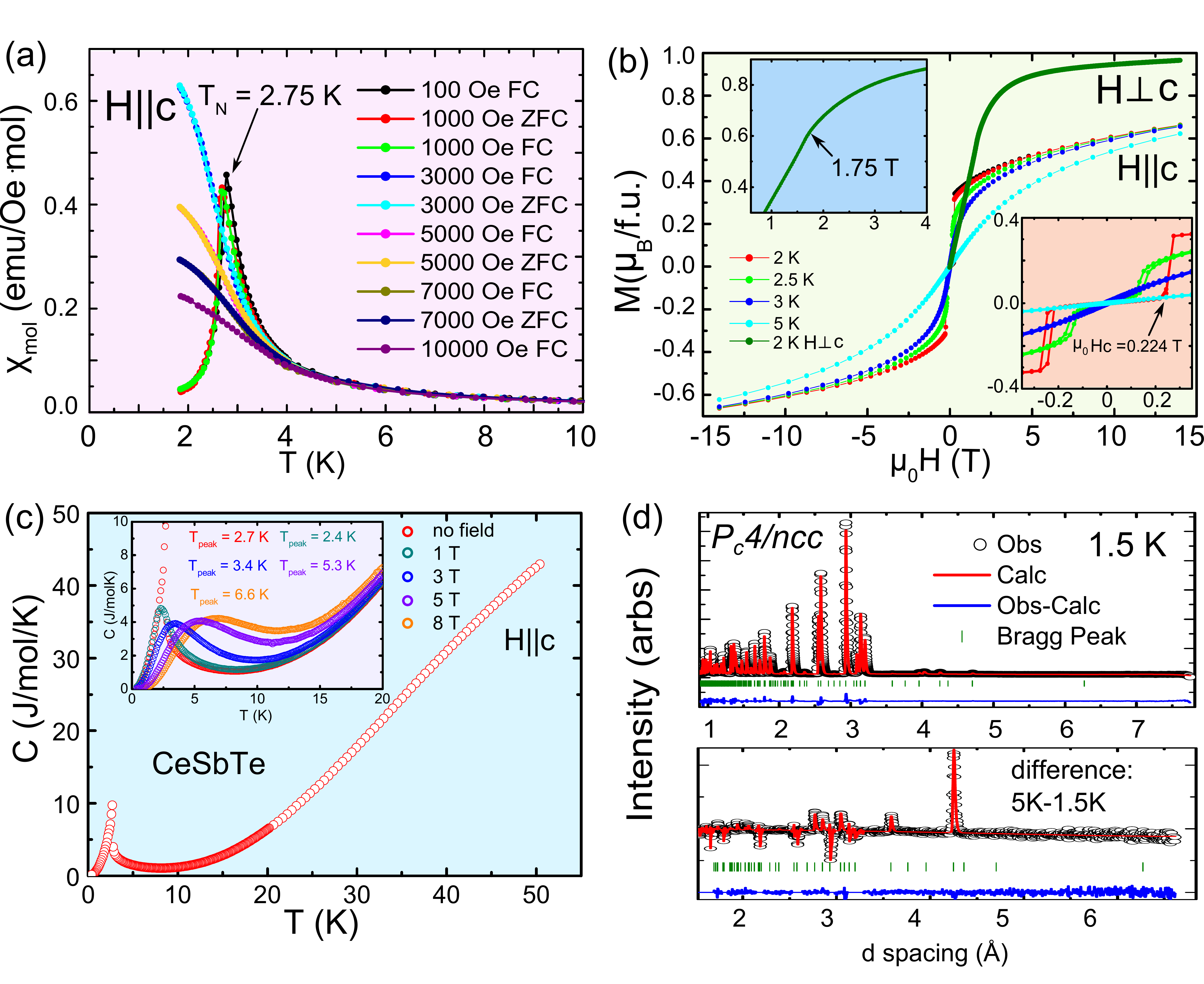}
  \caption{Magnetic Properties of $\rm CeSbTe$. (a) Temperature dependent magnetic susceptibility; different colored lines represent different applied field strengths ($H\parallel c$). (b) Field dependent magnetic susceptibility; different colors represent different temperatures. Below $T_\mathrm{N}$, a field dependent magnetic transition is observed that is reached at lower field strength for $H\parallel c$, but the saturation moment is reached faster with $H\perp c$. (c) Specific heat of $\rm CeSbTe$. The magnetic transition is clearly visible at 2.7\,K. The inset shows the behavior with different applied field strengths. (d) Refinement of neutron diffraction data taken at 1.5\,K, below the magnetic transition. The lower panel shows the pure magnetic diffraction pattern, which was obtained by subtracting the 5\,K data from the 1.5\,K data. The flat region in the difference line corresponds to regions excluded from the refinement, due to the presence of a small unknown impurity.}
  \label{mag}
\end{figure*}

Temperature dependent measurements of the magnetic susceptibility of a single crystal of $\rm CeSbTe$ are shown in figure \ref{mag}(a). The field was applied along the \textit{c}-axis (see SI for the other field orientation). For data with low field strengths the susceptibility shows a sharp cusp at the N\'{e}el temperature $T_\mathrm{N} = 2.75\,$K, indicative of the emergence of AFM order below this temperature. For field strengths larger than 0.25\,T the cusp disappears and the susceptibility keeps increasing with decreasing temperature, saturating at very low temperatures, reminiscent of FM behavior. Curie-Weiss fits (see SI) reveal an effective moment of 2.50\,$\mu_\mathrm{B}$ per formula unit which matches the expected free ion value of 2.54\,$\mu_\mathrm{B}$ for a Ce$^{3+}$ ion well. Deviations from the Curie-Weiss law at low temperatures are due to crystal electric field (CEF) splitting of the six-fold degenerate ground state of Ce$^{3+}$. Fits of the susceptibilities indicate a separation of the first excited CEF doublet of $\approx$150\,K. The field dependent data (Fig.\,\ref{mag}(b)) elucidate the transition from AFM to FM phase; at a field of $\mu_0H_\mathrm{c} = 0.224\,$T, the field dependent susceptibility shows a sudden increase. As shown in Fig.\,\ref{mag}(a) the required field is higher (1.75\,T) if the field is applied perpendicular to the \textit{c}-axis, and the jump is less pronounced. This indicates that the moments prefer to align along the \textit{c}-axis (easy axis). Interestingly, the expected saturation moment of a Ce$^{3+}$ ion (1\,$\mu_\mathrm{B}$ per Ce) is reached faster if the field is aligned perpendicular to the \textit{c}-axis which indicates an easy plane for the magnetization.
Similar field dependent magnetic transitions have also been observed in different square lattice Ce compounds, such as $\rm CeTe_2$ or $\rm CeSbSe$ \cite{park1997influence,jung2000competing,stowe2000crystal,chen2017devil}.

The magnetic transition was additionally investigated with specific heat measurements (see Fig. \ref{mag}(c)). At zero field, a magnetic transition is clearly visible at 2.7\,K, agreeing well with the susceptibility data. While the transition is shifted to slightly lower temperatures with a small applied field (here the field is applied along the \textit{c}-axis), which is typical for antiferromagnetic order, it shifts to higher temperatures when the applied field is larger that 1\,T, as expected for ferromagnetic order. The crossover between the two different types of magnetic order is therefore also reflected in the specific heat. For information about the magnetic entropy, see SI.

\begin{figure}[h]
  \centering
  \includegraphics[width = \columnwidth]{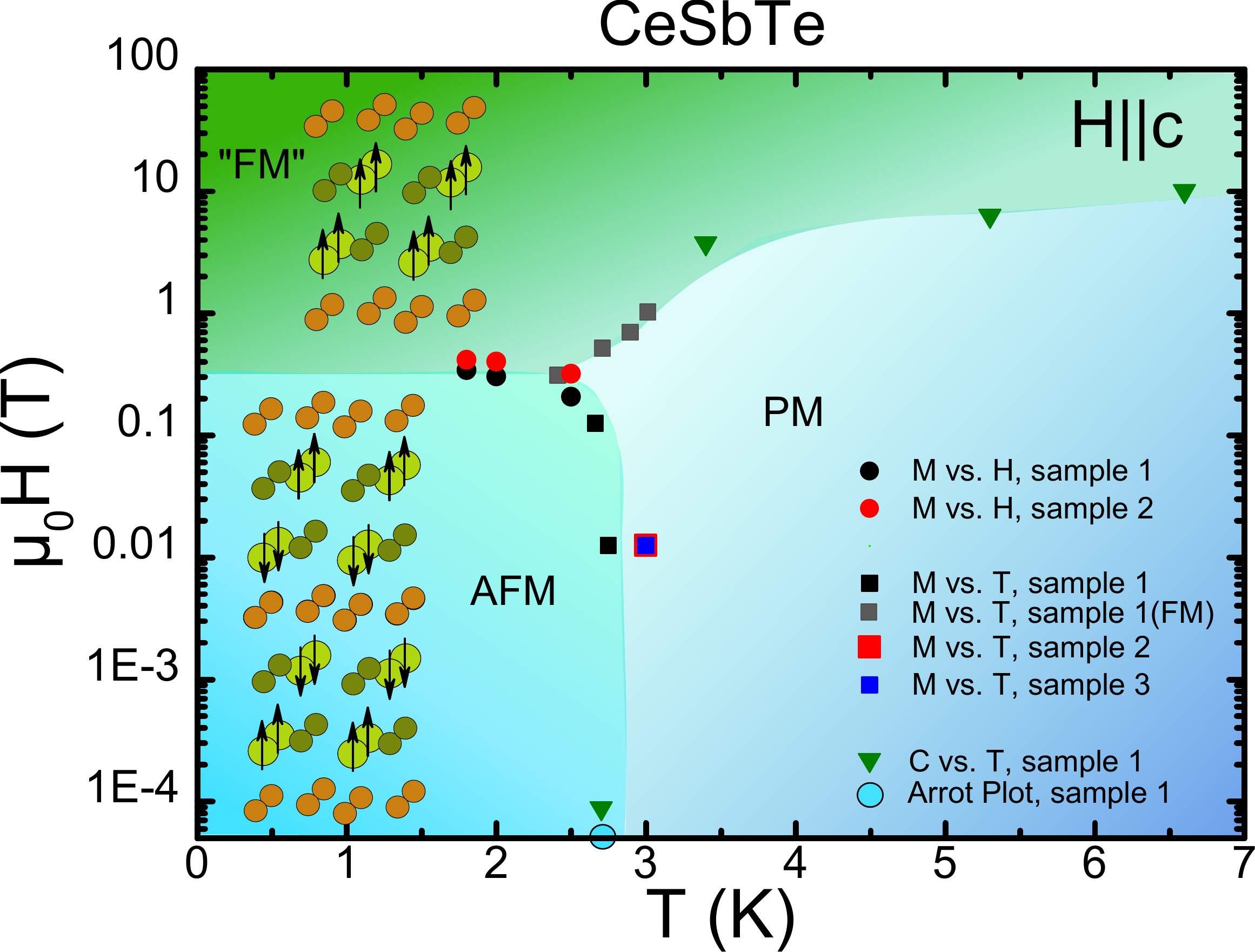}
  \caption{Magnetic phase diagram of $\rm CeSbTe$.  Three different magnetic phases can be observed within a low field limit. The magnetic structures of the different phases are shown in the respective regions. Different colored dot shapes belong to different samples and different measurement techniques, respectively.}
  \label{pd}
\end{figure}

To analyze the low field magnetic structure of $\rm CeSbTe$, neutron powder diffraction measurements were performed. Some refinements are shown in Fig.\,\ref{mag}(d), and more detailed figures and tables with refined parameters can be found in the SI. As mentioned before, at 5\,K, above the N\'{e}el temperature, the refinement yields the tetragonal space group \textit{P4/nmm}, in agreement with the results form SXRD and PED. Upon cooling, additional Bragg peaks appear, due to magnetic order. These extra reflections can be indexed with a propagation vector $\boldsymbol{k}=(0,0,\frac{1}{2})$, and the absence of the (00\textit{l}$\pm\frac{1}{2}$) reflections clearly indicates that the spins are oriented along the \textit{c} direction, matching the indications from the susceptibility data (see SI for details). The magnetic structure was solved at 1.5\,K with a doubling of the unit cell along \textit{c} in the magnetic space group $P_c4/ncc$ (No.\,130.432, type IV) corresponding to the $mZ_1^+$ irreducible representation (see Fig.\,\ref{mag}(d) for a fit of the nuclear and magnetic contributions as well as just the pure magnetic diffraction pattern). A drawing of the magnetic structure can be found in Fig.\,\ref{pd}. In each Ce single layer the spins are ferromagnetically arranged and coupled antiferromagnetically with the following layer, forming an antiferromagnetic bilayer. The bilayers are then coupled antiferromagnetically with each other, which causes the doubling of the unit cell along \textit{c}. Further information regarding the magnetic structure solution is given in the SI.

The magnetic phase diagram of $\rm CeSbTe$ is summarized in Fig.\,\ref{pd}, which includes the results of the susceptibility, specific heat and neutron diffraction measurements, measured on multiple and distinct single crystals. In order to find the hypothetical zero field magnetic transition temperature from the susceptibility data we used the Arrott plot analysis (see SI for additional information). The transition temperatures in the susceptibility data were inferred from the derivative $\mathrm{d}\chi/dT$. There are three regions in the phase diagram; at temperatures below $T_\mathrm{N} = 2.75\,$K and $\mu_0H_\mathrm{c} = 0.224\,$T the spins order antiferromagnetically, above the critical field, they switch to a ferromagnetic type order, and above $T_\mathrm{N}$ and low field, the material is paramagnetic. Note that the spin flop transition to the ferromagnetic phase happens at higher field strength if the field is aligned perpendicular to the \textit{c}-axis. Depending on the field alignment the spins can be adjusted to point in different crystallographic directions. 

\section{Discussion}

\begin{figure*}[h]
  \centering
  \includegraphics[width = \textwidth]{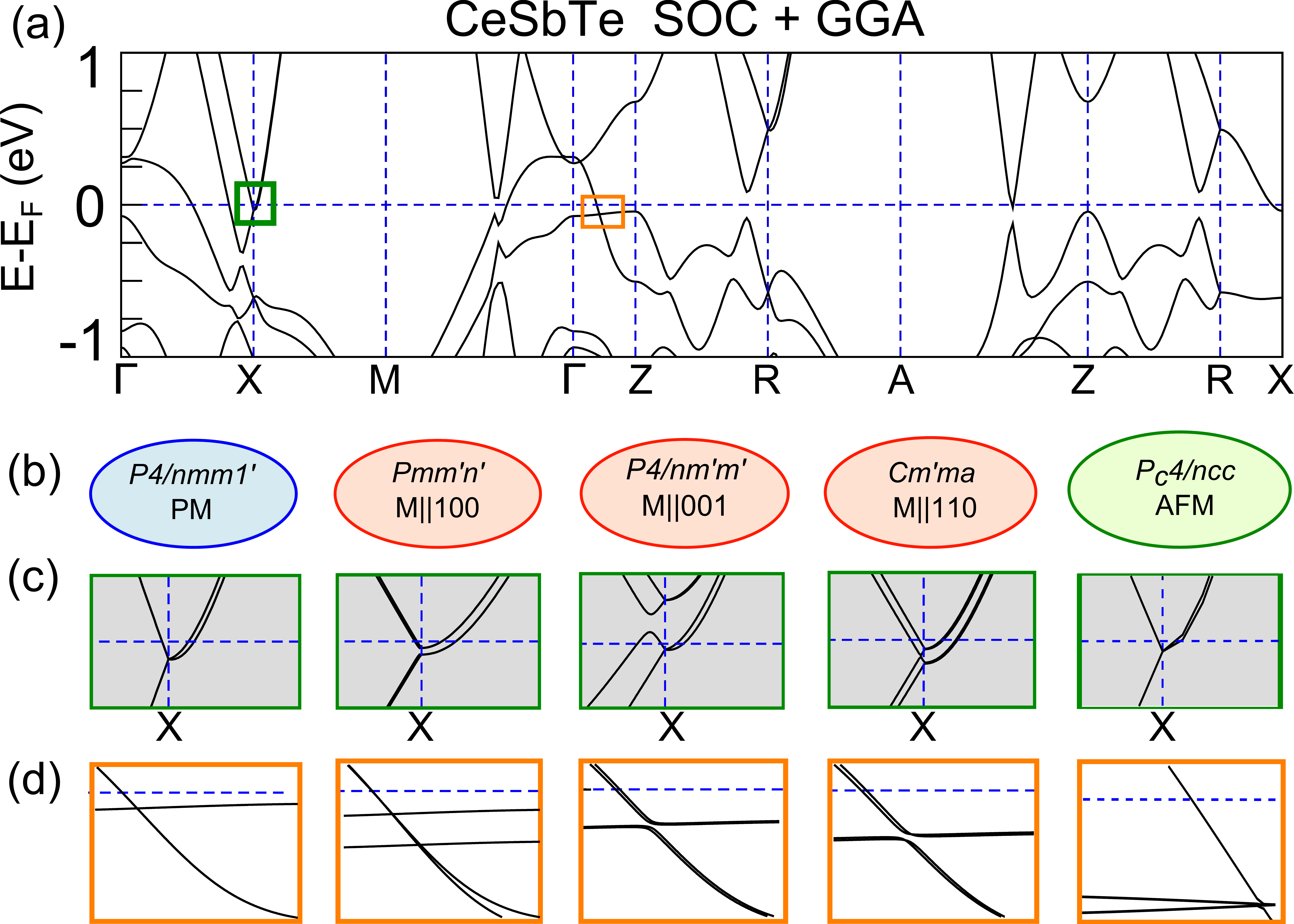}
  \caption{Band structure of $\rm CeSbTe$. (a) Calculated paramagnetic band structure including SOC. The green and orange box highlight the nonsymmorphic Dirac and the tilted Dirac crossing, respectively. (b) Symmetry groups of the different accessible phases. (For drawings of the respective magnetic structures see Fig.\,12 in the SI) (c) Detailed plots of the region around the X point and how different types of magnetic order affect the electronic structure. (d) Detailed plots of the region along $\Gamma$Z and the affect of magnetic order.}
  \label{calcs}
\end{figure*}

The different possibilities of magnetic order are expected to have a significant effect on the electronic structure of $\rm CeSbTe$. In the paramagnetic case, TRS is preserved, conserving the four-fold degeneracies at the nonsymmorphically protected high symmetry points. In the antiferromagnetic and ferromagnetic phases it is broken, which likely leads to a lifting of degeneracies. Figure \ref{calcs}(a) shows the calculated band structure of the paramagnetic phase of $\rm CeSbTe$. Without consideration of \textit{f}-electrons and spin-polarization, four-fold degenerate lines, protected by joint IS, TRS, and nonsymmorphic symmetries exist on the boundary of the BZ, such as that along RX, similar as in $\rm ZrSiS$ and related compounds \cite{chen2017dirac}. At the X point, one of these crossings appears very close to the Fermi level (see green highlight). Further four-fold degeneracies appear at M and A, but those are much further away from the Fermi level. Along $\Gamma$Z there is a slightly tilted Dirac crossing that is protected by the four-fold rotation symmetry in the tetragonal space group \textit{P4/nmm}.

If magnetism is included in the calculation, the electronic structure is affected rather severely dependening on the type of magnetic order and the orientation of the spins. As long as spin-polarization is taken into consideration, the symmetry is reduced by breaking TRS, leading to splitting of the four-fold degenerate lines. Fig.\,\ref{calcs}(c) shows the region around the X point, when different types of spin polarizations are included in the calculation. For complete plots of the electronic structure including spin polarization, see SI. Without SOC, the nonsymmorphic symmetry $\overline{C}_{2x} = \{C_{2x}\vert(\frac{1}{2},0,0)\}$ guarantees the double degeneracy at X point. With SOC, each band remains two-fold degenerate, since $(TI)^2 = -1$, where T is time-reversal symmetry and I is the spatial inversion symmetry. The screw $\overline{C}_{2x}$ anticommutes with I at the X point, leading to a four-fold degeneracy \cite{young2015dirac}. When a ferromagnetic spin polarization is taken into account, time reversal symmetry is broken and the Dirac point at X splits correspondingly. The details of the band structure are strongly dependent on the orientation of the magnetization. If the spins are aligned along [001], only TRS is broken, while the anti-unitary combination of the nonsymmorphic symmetry $\overline{C}_{2x}$ and time-reversal $\overline{C}_{2x}T$ is preserved. Therefore, the four-fold degeneracy is reduced to a double degeneracy at X as well as doubly degenerate bands for all $\boldsymbol{k} = (\pm \pi, k_y,k_z)$. Note that this scenario appears below 2.75\,K and $\mu_0H_\mathrm{c} = 0.224\,$T, if the field is aligned along the crystallographic \textit{c} axis. If the spins are aligned along [100], only the unitary nonsymmorphic symmetry $\overline{C}_{2x}$ remains. A new anti-unitary symmetry M$_z$T (where M$_z$ is a mirror operation) is present additionally, protecting a double degeneracy along the line XR. The same symmetry appears for a FM spin orientation along the [110] direction. In the AFM phase, even if TRS is not a symmetry of the Schr\"{o}dinger Hamiltonian anymore, the symmetry element $ \{\bar 1\vert(0,0,\frac{1}{2})\}$ squares to $-1$ and maps \textit{k} to \textit{k}, such as TRS would do in space group 130. This is a result from the doubled \textit{c}-axis in this phase. Therefore, although $\rm CeSbTe$ crystallizes in space group 129, its AFM phase behaves like a material in space group 130 with TRS. This causes the bands to be four-fold degenerate at  X, R and M and eight-fold at A at $E_i=-4.6$\,eV, see Fig.\,13 in the SI). We can conclude that AFM order is a possibility to create higher order degeneracies that are associated with new types of topological phases in space groups that have previously not been considered. This considerably extends the amount of materials that can exhibit such unconventional quasiparticles.  

The changes in symmetry also affect the electronic structure away from high symmetry points. For example, two-fold degenerate Weyl points appear close to the X point if the spins are aligned along [001]. Additionally, there is an effect on the tilted four-fold degenerate Dirac crossing along the $\Gamma$Z direction (Fig.\,\ref{calcs}(d)). In the ferromagetic phase, with spins aligned along the [001], one of these bands splits, whereas the other one remains degenerate, resulting in a triply degenerate point close to the Fermi level that is composed of two different irreducible representations of 2+1 dimensions along the line $\Lambda$ connecting $\Gamma$ and Z. Triply degenerate band crossings have also been associated with the possibility to create double Fermi arcs at the surface \cite{bradlyn2016beyond}.

$\rm CeSbTe$ is thus a unique material that allows access to many different magnetic groups that allow for irreducible representations in various dimensionalities. Fig.\,12 in the SI summarizes the accessible magnetic groups and how the system can be transformed to exhibit each symmetry. Drawings of the crystal structure of $\rm CeSbTe$ in each magnetic group are also shown in the same figure. The different dimensions of the irreducible matrix representations at each time reversal inversion momentum point are also given in the SI. The magnetic groups that are accessible in $\rm CeSbTe$ can either be symmorphic or nonsymmorphic, which adds to the richness of different irreducible representations accessible.

\section{Conclusion}
In conclusion, we showed that the tetragonal compound $\rm CeSbTe$ can host many different topological features in its electronic structure, ranging from four-fold and two-fold Dirac and Weyl crossings to more exotic eight-fold and three-fold degeneracies. Due to the easy access to several different magnetic phases, this system is a rich playground to study the effects of symmetry breaking due to magnetic order on different types of Dirac crossings, as well as on higher order degeneracies that result from nonsymmorphic symmetry. In particular, the AFM phase is represented by a space group of higher symmetry order, the nonsymmorphic group $P_c4/ncc$. Due to the added symmetry elements, the material shows the same irreducible representations as materials that crystallize in space group 130 with TRS. As a consequence, $\rm CeSbTe$ is  a unique nonsymmorphic magnetic topological material that displays exotic states such as eight-fold new fermions. This concept can also be extended to further materials crystallizing in space group 129 (a much more abundant space group compared to 130), significantly increasing the available materials that could host new fermions. Due to the difficulty of taking ARPES spectra in the presence of a magnetic field, future investigations of $\rm CeSbTe$ with scanning tunneling microscopy (STM) are of high interest. 

\section{Acknowledgements}
We gratefully acknowledge the financial support by the Max Planck Society, the Nanosystems Initiative Munich (NIM) and the Center for Nanoscience (CeNS). Leslie M. Schoop gratefully acknowledges financial support by the Minerva fast track fellowship. We also would like to thank Roland Eger for single crystal diffraction measurements.
This work was partially supported by the DFG within the proposal “Dirac materials in square lattice compounds” under proposal SCHO 1730/1-1. The authors acknowledge the science and technology facility council (STFC) for the provision on neutron beamtime at the ISIS facility (UK).

\end{document}